\documentclass[useAMS,usenatbib]{mn2e}

\usepackage{graphicx,rotating,longtable,lscape,color}

\def\lsim{\mathrel{\rlap{\lower 3pt \hbox{$\sim$}} \raise 2.0pt \hbox{$<$}}}
\def\gsim{\mathrel{\rlap{\lower 3pt \hbox{$\sim$}} \raise 2.0pt \hbox{$>$}}}

\title[The field of SDSS J0927]{A photometric study of the field around the
  candidate recoiling/binary black hole SDSS J092712.65+294344.0}
\author[Decarli, Reynolds \& Dotti]{R. Decarli$^{1}$\thanks{E-mail:
    roberto.decarli@mib.infn.it}, M. T. Reynolds$^{2}$ and M. Dotti$^{2}$\\ $^{1}$ Universit\`{a} degli Studi dell'Insubria, via
  Valleggio 11, 22100 Como, Italy\\ $^{2}$ Dept. of Astronomy, University of
  Michigan, Ann Arbor, MI 48109, USA }
\begin{document}

\date{ }

\pagerange{\pageref{firstpage}--\pageref{lastpage}} \pubyear{2009}

\maketitle

\label{firstpage}

\begin{abstract}
  We present a photometric FUV to K$_s$-band study of the field around
  quasar SDSS J092712.65+294344.0. The SDSS spectrum of this object shows
  various emission lines with two distinct redshifts, at $z=0.699$ and
  $z=0.712$. Because of this peculiar spectroscopic feature this source has
  been proposed as a candidate recoiling or binary black hole.  A third
  alternative model involves two galaxies moving in the centre of a rich
  galaxy cluster.  Here we present a study addressing the possible presence
  of such a rich cluster of galaxies in the SDSS J092712.65+294344.0 field.  We
  observed the $3.6\times2.6$ square arcmin field in the K$_s$-band and
  matched the NIR data with the FUV and NUV images in the GALEX archive and
  the $ugriz$ observations in the SDSS. From various colour-colour diagrams
  we were able to classify the nature of 32 sources, only 6--11 of which
  have colours consistent with galaxies at $z\approx0.7$.  We compare these
  numbers with the surface density of galaxies, stars \& quasars, and the
  expectations for typical galaxy clusters both at low and high
    redshift.  Our study shows that the galaxy cluster scenario is in clear
  disagreement with the new observations.
\end{abstract}


\section{Introduction}

SDSS J092712.65+294344.0 \citep[hereafter S0927;][]{adelman08}, is a quasar
with a very peculiar spectrum. It exhibits a set of optical broad and narrow
emission lines (``b-system''), blueshifted $\approx 2650$ km s$^{-1}$ with
respect to a second set of narrow emission lines (``r-system'').

Recently three different models have been proposed in order to explain the
peculiar spectral features of S0927.  Based on the recent results of
\citet{schnittman07} and \citet{campanelli07}, \citet{komossa08} suggest
that S0927 is the first candidate recoiling remnant of a massive black hole
(MBH) binary coalescence, ejected from the nucleus of its host galaxy by
gravitational radiation recoil.  In this model the (broad and narrow)
b-system lines are emitted by gas comoving with the recoiling MBH, while the
r-system lines are emitted by the gas in the host galaxy.

\citet{bogdanovic09} and \citet{dotti09} noticed that Komossa's model
depends on an unlikely combination of parameters for the coalescing MBH
binary, and has difficulty explaining the narrow emission lines of the
b-system. These two papers proposed an alternative model, assuming the
presence of a sub--parsec separation MBH binary in the centre of the host
galaxy. In this model the blueshift of the b-system is related to the
orbital motion of the MBH binary, while the r-system corresponds to the
narrow line region of the host.

A third model has been discussed in \citet{heckman09} and
\citet{shields09}. The authors consider the possibility of a chance
superposition of two galaxies, assuming the presence of a rich galaxy
cluster hosting S0927. The different systems of lines at different redshifts
are produced in two different galaxies, moving at high relative velocity
deeply inside the potential well of the cluster. This model is the simplest
explanation for the three sets of lines, and is immediately testable\footnote{For the other two models, an immediate test is more
  difficult. The binary model predicts the redshift of the b-systems
  of emission lines to change on a time-scale of tens of years. On the other
  hand, the ejection hypothesis does not have any falsifiable prediction if
  the MBH is recoiling along the line of sight. This is the most plausible
  configuration: any other configuration would imply higher, extremely
  improbable, de--projected recoiling velocities. However, if the recoiling
  MBH also has a large velocity component in the plane of the sky, the
  quasar could be displaced from the centre of the host galaxy. Such an
  offset can be observed with the Hubble Space Telescope.}
Such a high velocity difference between the two galaxies is inconsistent
with a simple on--going merger, and requires the deep potential well of a
rich galaxy cluster. From the study of SDSS $ugriz$ photometry,
\citet{heckman09} report the detection of a number ($\sim10$) of faint, red
sources within a $4 \times 4$ square arcmin box around S0927, which
are consistent with early-type galaxies at $z\approx0.65$. The analysis in
\citet{heckman09} is limited by the fact that the optical colours of red
stars and of galaxies at $z=0.7$ are similar, and the latter are practically
unresolved under typical SDSS seeing conditions. To improve this study, we
obtained a deep NIR image of the field of S0927. Moreover, following
\citet{niemack09} we also consider the UV information from GALEX to better
assess the nature and redshift of field sources. We will then compare the
number of sources found consistent with $z=0.7$ with the expectations for a
galaxy cluster at that redshift.

Throughout the paper, we will adopt a concordance cosmology
with $H_0=70$ km s$^{-1}$~Mpc$^{-1}$, $\Omega_m=0.3$, $\Omega_\Lambda=0.7$.
Within this cosmology, the distance modulus at $z=0.7$ is $43.15$ mag and 
the angular scale is $7.15$ kpc/$''$.

\section[]{NIR observations and data reduction}

\begin{figure*}
\begin{center}
\includegraphics[width=\textwidth]{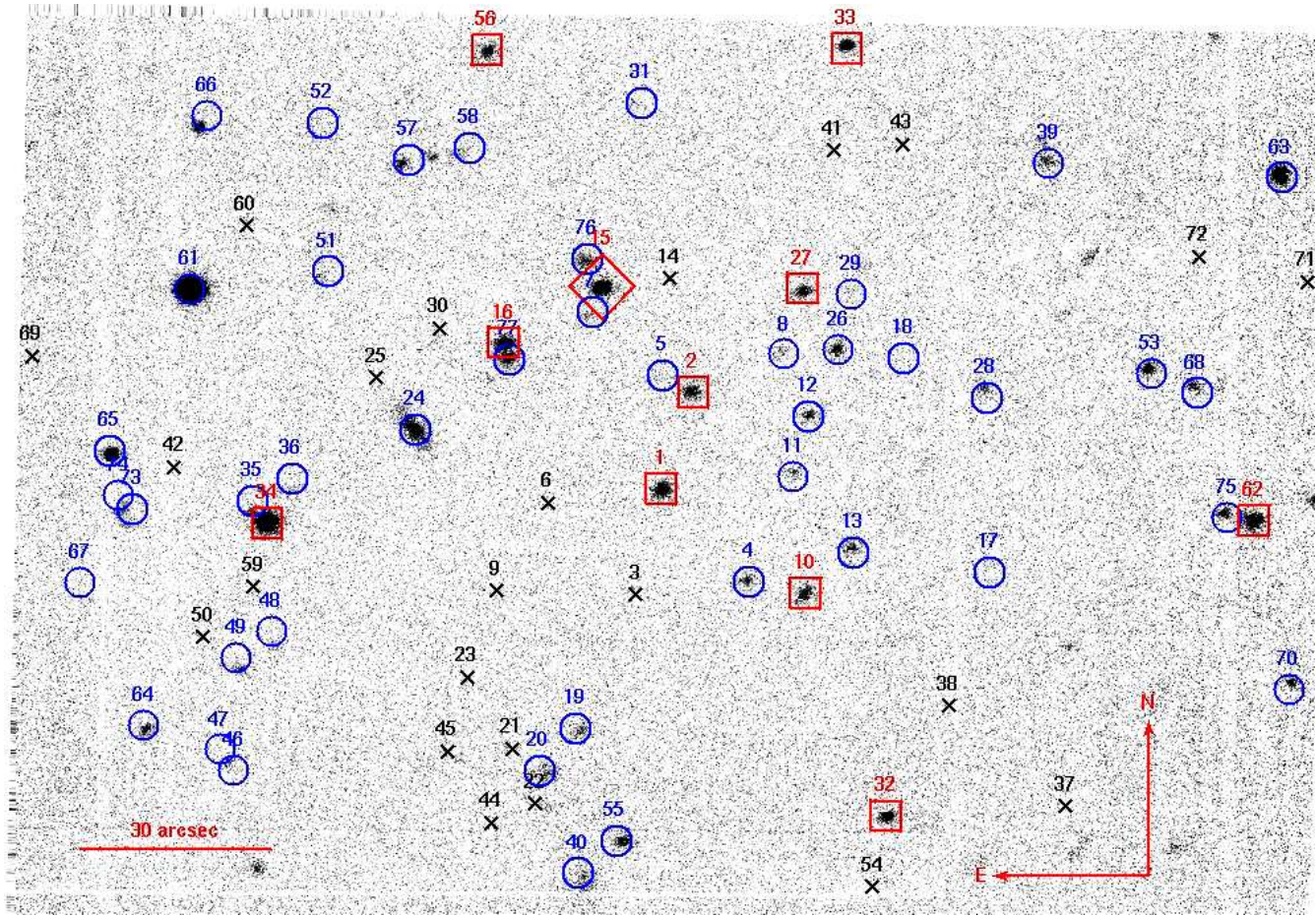}\\
\caption{The K$_s$-band image of the field around S0927, which
is labelled with n.15 (diamond). Objects labelled from 1 to 74 are 
present in the SDSS photometric catalogue, the others are reported here for 
the first time. Crosses mark the sources below our sensitivity limit in 
all the available images. Squares refer to the 11 objects consistent 
with $z\approx0.7$ galaxies, according to our colour-colour diagnostics. All of
the other sources are labelled with circles.
}\label{fig_detK}
\end{center}
\end{figure*}

K$_s$-band photometry of the S0927 field was obtained using TIFKAM
\citep{pogge98} at the MDM Observatory 2.4m telescope on the night of 2008
November 3. The observing conditions were good with a seeing of
$1.0''$ -- $1.2''$ throughout. The individual exposure times were 20s,
with 5 coadds per image, resulting in 100s integration time per image. The
telescope was dithered using a $3\times3$ grid to allow for accurate
subtraction of the NIR background.  The data were processed using standard
\textsc{iraf} routines\footnote{\textsc{iraf} is distributed by the National
  Optical Astronomy Observatories, which are operated by the Association of
  Universities for Research in Astronomy Inc., under a cooperative agreement
  with the National Science Foundation.}.  The final mosaiced image,
displayed in Fig. \ref{fig_detK}, was created using the \textsc{xdimsum}
package. The total exposure time is 90 minutes.

Astrometric calibration was performed through comparison with the USNO
database. The photometric zero point (ZP) was computed by comparing the
instrumental magnitudes of field stars to the 2MASS catalogue (see Table
\ref{tab_bands}). From the RMS of sky counts within a seeing radius, we
estimate a 3-$\sigma$ limiting magnitude of $m_{\rm Ks}^{\rm
  limit}=19.45$.

\section[]{Archive data}
Images of the field of S0927 are publicly available in the archives of the
Galaxy Evolution Explorer (GALEX) and of the Sloan Digital Sky Survey
(SDSS).  

GALEX features two broad band filters at central wavelengths of
$\sim1500$\AA~(FUV) and $\sim2300$\AA~(NUV). The field of S0927 was imaged
as a part of the GALEX All Sky Survey. The point spread function (PSF) in
the FUV and NUV images is $3.7''$ and $5.3''$, respectively, with a pixel
scale of $1.5''$/pix. The exposure time of the two images is
112s. Assuming the zero points provided in the GALEX website (cfr Table
\ref{tab_bands}\footnote{
    \texttt{galexgi.gsfc.nasa.gov/docs/galex/FAQ/counts\_background.html};
    see also \citet{morrissey05}.}), this yields a limiting magnitude
of $m_{\rm FUV}^{\rm limit}=19.9$ and $m_{\rm NUV}^{\rm limit}=20.2$.

The SDSS Data Release 6 \citep{adelman08} provides $ugriz$ photometry of
nearly a quarter of the sky. The typical seeing is $\sim 1.5''$. The pixel
scale is $0.396''$/pix. We estimate the photometric zero points of each band
by a comparison with the instrumental magnitudes of the sources and the
values reported in the SDSS catalogue.  The zero points, magnitude limits
and angular resolution of the available images are summarised in Table
\ref{tab_bands}.

\begin{table}
\begin{center}
\caption{Properties of the available images. (1): filter. (2) zero point.
  We list the mean and RMS values, estimated through the comparison 
  between catalogue and instrumental magnitudes of field sources. (3) 
  limiting magnitude due to background counts. (4) pixel scale. (5) point 
  spread function (=seeing in ground based observations). (6) number of 
  sources exceeding the sensitivity limit in each band.} \label{tab_bands}
\begin{tabular}{cccccc}
   \hline
   Filter & ZP             & Limit mag & Pixel scale & PSF  & N.det. \\
          & [mag]          & [mag]     & [$''$/pix]  &[$''$]&        \\
   (1)    &  (2)           &  (3)      &  (4)        & (5)  & (6)    \\
   \hline
    FUV   & $18.82$        & $19.9$    & $1.5$       & $3.75$ &  3 \\
    NUV   & $20.08$        & $20.2$    & $1.5$       & $5.30$ &  3 \\
    $u$   & $23.32\pm0.25$ & $21.24$   & $0.396$     & $1.18$ &  5 \\
    $g$   & $24.17\pm0.17$ & $22.17$   & $0.396$     & $1.11$ & 22 \\
    $r$   & $23.93\pm0.09$ & $21.72$   & $0.396$     & $1.07$ & 29 \\
    $i$   & $23.54\pm0.12$ & $21.15$   & $0.396$     & $0.94$ & 44 \\
    $z$   & $22.39\pm0.10$ & $20.07$   & $0.396$     & $0.94$ & 21 \\
    K$_s$ & $22.98\pm0.10$ & $19.45$   & $0.206$     & $0.92$ & 47 \\
   \hline
   \end{tabular}
   \end{center}
\end{table}

\section[]{Data analysis}
We limit our analysis to the $3.6\times2.6$ arcmin region around S0927
covered in all 8 bands (where the limit is provided by the region observed
in the K$_s$-band image).

We re-measured the magnitudes of each source in all the available
bands. We consider a source as detected if its flux (in counts) is
larger than three times the RMS of the sky in the PSF area. The
SDSS photometric catalogue lists 74 sources in our frame. Additional three
sources are detected in the K$_s$-band but do not appear in the SDSS
catalogue. As they lie close to bright companions, we argue that the
SDSS source selection algorithm fails to de-blend them. The whole
sample therefore consists of 77 objects.

 GALEX images are the shallowest: Only 3 sources are detected in FUV and
  NUV, and they all appear in the images in the other bands.  Six sources
  are detected in the SDSS but fall below our sensitivity limit in
  K$_s$. From visual inspection, we find that all of them are faint, early
  spectral-type stars which we miss in K$_s$-band due to their blue SED.

Our criterion is somewhat stricter than the SDSS detection algorithm,
hence 23 sources are not detected in any of our bands. From a careful
  inspection, we find that ten of them are extremely faint sources
  ($m\gsim24$ in the SDSS photometric catalogue) and are possibly artifacts
  of the SDSS source detection algorithm.  The remaining sources are
  detected with a lower significance (e.g., if we relax our threshold to
  2-$\sigma$ fluctuations over the sky count RMS in the PSF area, 6 other
  sources are detected). In section \ref{sec_cluster} we will see how
  relaxing our sensitivity limit would affect the results of our analysis.
The remaining 54 sources, detected in at least a single band, are listed
in Table \ref{tab_allsources}.

We note that at $z=0.7$, 5 kpc $\approx$ $0.7''$, thus most of the
light from galaxies at this redshift would be enclosed in the typical
seeing of our data. This is why morphology cannot help in the
identification of the nature of our sample sources: With the 
exception of sources n.24 and n.40, which are well-resolved foreground
galaxies, all the objects are practically unresolved.

\section[]{Classification of the observed sources}
We use several colour-colour diagnostic diagrams to select 
possible galaxy candidates. This can be applied on a 
limited number of sources, because of the relative shallowness of
the SDSS photometry in each particular band. Note that all
our colour cuts are intended to provide a rule of thumb for
distringuishing between stars and possible galaxy candidates.

We derive reference colours of galaxies and quasars by measuring the
spectral magnitudes in each band from the Elliptical and Sc galaxy templates
by \citet{mannucci01} and the quasar template by \citet{francis91},
redshifted to $z=0.7$.  We also show the locus of main sequence stars in our
diagrams using the work by \citet{girardi05} and the prediction of
\textsc{TRILEGAL}
software\footnote{\texttt{http://stev.oapd.inaf.it/cgi-bin/trilegal}} as
reference.

\begin{figure*}
\begin{center}
\includegraphics[width=0.89\textwidth]{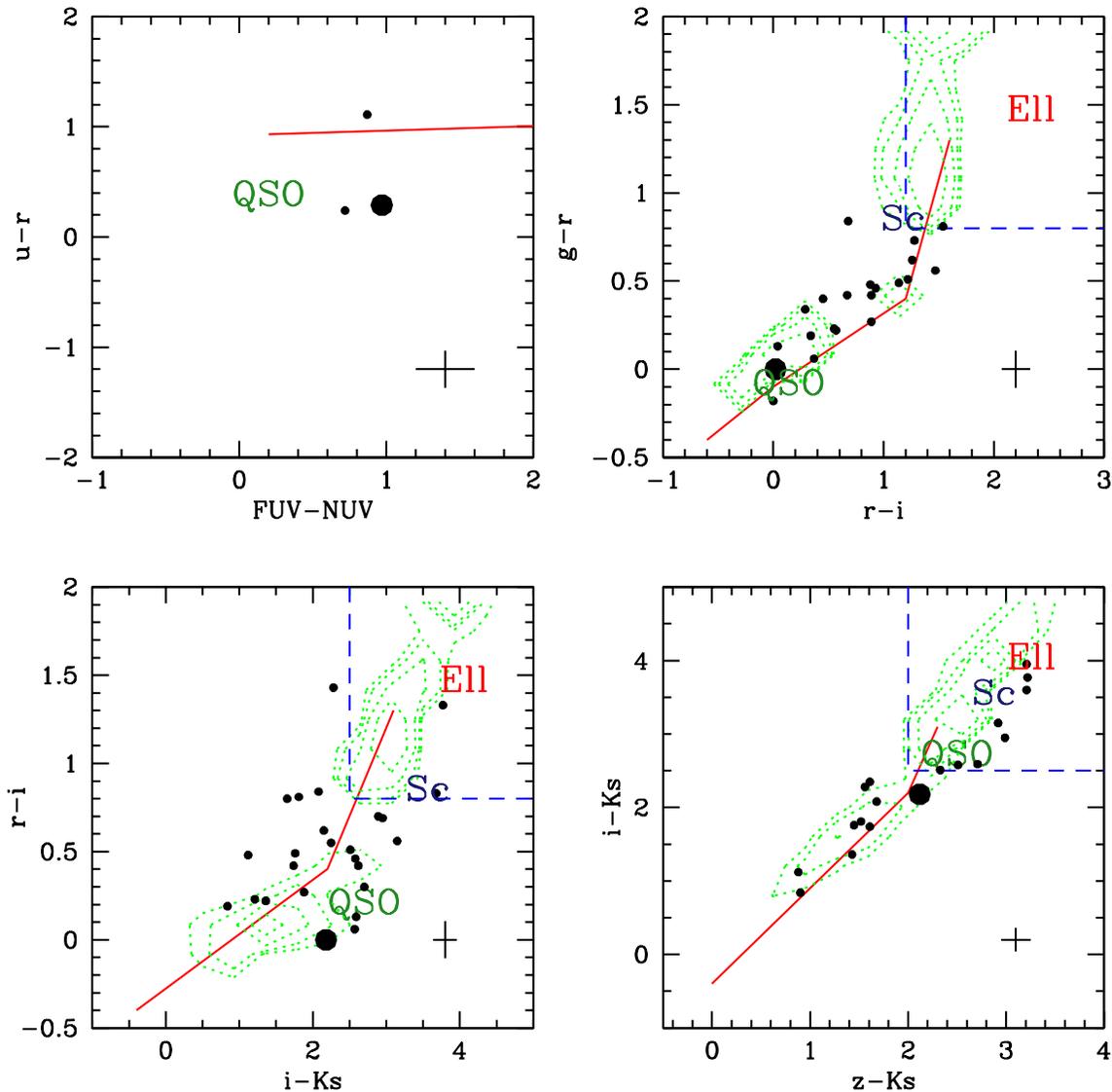}\\
\caption{Colour-colour diagrams used as diagnostic of the nature of the
  observed sources. Dots mark the sources in the present study. Note that
  the number of points necessarily changes from panel to panel, according to
  source detection rates in each band. The large dot refers to
  S0927. Comparison sources are quasar (QSO), elliptical (Ell) and spiral
  (Sc) galaxy templates redshifted to $z=0.7$ and the loci of main sequence
  stars according to \citet{girardi05} (solid line). The dotted contours
  refer to the distribution of spectroscopically confirmed stars in the
  SDSS+2MASS, in a $5\times5$ square degrees area around S0927 (note that
  the SDSS spectroscopic survey is inefficient at detecting main
  sequence stars, hence this comparison is merely qualitative of the scatter
  around Girardi's model).  The colour cuts described in the text are shown
  with dashed boxes.  Typical error bars on the colours are also indicated
  in the bottom-right corner of each panel.}\label{fig_colors}
\end{center}
\end{figure*} 

The (FUV-NUV,$u$-$r$) plane allows us to infer the nature of the bluest
sources. Due to the shallowness of the All Sky Survey, only 3 sources are
detected by GALEX: n.15, n.16 and n.40. The first two are S0927 and a
companion source $\sim0.5$ mag fainter in all the observed bands. We suggest
that this is a quasar with a redshift close to $0.7$. Source n.40 is a
clearly resolved low-redshift irregular galaxy, the stellar population of
which is dominated by young stars, as confirmed by our plot.

The ($g$-$r$,$r$-$i$) diagram shows that the sources detected in all these
bands are consistent with Galactic stars following the Main
Sequence. Only the reddest sources ($g$-$r>1.2$ and $r$-$i>0.8$) are 
marginally consistent with the Sc template at $z=0.7$
\citep[e.g.][]{csabai03}.

The contribution of NIR data is evident in the ($i$-K$_s$,$r$-$i$) and
($z$-K$_s$,$i$-K$_s$) diagrams, where many more red sources are
detected. Again, only the reddest sources ($i$-K$_s$$>2.5$ and
$z$-K$_s$$>2$) are consistent with galaxies at $z=0.7$. Colour cuts are a
simplified adaption of those adopted in \citet{blanc08} (where we adopt
$i^{\rm AB}\approx i^{\rm SDSS}$, $z^{\rm AB}\approx z^{\rm SDSS}$ K$_s^{\rm
  AB}$=K$_s^{\rm MDM}$+$1.9$)\footnote{See
  \texttt{http://www.sdss.org/dr5/algorithms/fluxcal.html} and
  \texttt{http://www.eso.org/science/goods/releases/20050930/}}.

All the adopted colour cuts favour false-positive detections, and heavy
contamination from very red stars is expected. As a consequence, the
number of $z \approx 0.7$ galaxy candidates found has to be considered an
upper limit.

In Table \ref{tab_classif}, we list the 32 sources for which at least two
colours are available. In addition to the quasar hosts n.15 (S0927) and
  n.16, four sources (n.1, 2, 10, 27) satisfy all the colour selections we
  set for $z\approx0.7$ galaxies, in the bands they are detected. A further
three sources (n.34, n.56 and n.62) fulfill all but one constraint. Finally,
n.32 and n.33 are consistent with 2 out of 4 colour conditions. The number
of galaxy candidates at $z\approx0.7$ is thus 6 (4 quiescent galaxies + 2
quasar hosts), with at most 5 extra `lower quality' candidates. On the
  other hand, our colour-colour diagrams show that 21 out of 77 sources have
  colours inconsistent with galaxies at $z=0.7$. We argue that they are
  Galactic stars or lower redshift galaxies.

\citet{heckman09} find 10 galaxy candidates South-West of S0927. 
Indeed, 6 out of 11 candidates from our analysis are found in the South-West
quadrant. Though, since \citet{heckman09} do not provide any other 
indication on the position of their candidates, we cannot assess whether
they match ours or not.

\section[]{Comparison with expectations}
In order to evaluate whether the number of sources we find is consistent
with the galaxy cluster scenario, we must take into account the expected
number of contaminating detections in the field of S0927, i.e. field
galaxies, stars and quasars. Then we will compare the remaining source
counts with the expectations for a typical galaxy cluster at $z\approx0.7$,
given our flux limit. We will focus on the detections from the K$_s$ image,
which is much deeper than other available data and probes the rest-frame
J-band of galaxies, i.e. is almost insensitive to the age of the stellar
populations.  Furthermore, given the shallowness of the SDSS images and
  the red colours of galaxies at $z=0.7$, only the brightest galaxies in the
  observed K$_s$-band luminosity function (K$_s\lsim17$, see the colour cuts in
  Figure \ref{fig_colors} and the limiting magnitudes in Table
  \ref{tab_bands}) can also be detected in the available optical images.

\subsection{Field galaxy, star and quasar surface densities}\label{field_stuff}
\begin{description}
\item[{\it i- Galaxies:}] We first compare with the average surface density
  of galaxies and stars from the general field, following the approach
  presented in \citet{fukugita04}. From the MUSYC survey, \citet{blanc08}
  evaluate the galaxy number counts per square degrees in the K$_s$-band.
  Considering all the sources flagged as `pBzK' or `sBzK' in the MUSYC
  survey, that is all {\it bona fide} galaxies, and dropping all the objects
  fainter than our K$_s$-band sensitivity limit, the expected surface
  density is $\approx2.7$ galaxies per square arcmin, that is, $\sim 25$
  galaxies in our field.  The galaxy surface density estimate provided by
  \citet{blanc08} has a 10 per cent uncertainty, accounting for cosmic
  variance and Poissonian errors.
\item[{\it ii- Stars:}] Through the \textsc{TRILEGAL} software, we also
  estimate the expected number of Galactic stars in the direction of
  S0927. Assuming a Chabrier log-normal initial mass function and a
  thin-disc model of the Galaxy with 2.8 kpc of scale radius \citep[see][for
    details]{girardi05}, we estimate that the expected number of stars with
  K$_s$$<19.45$ in our frame is $\sim 11$. Adopting different assumptions in
  terms of stellar initial mass function and Galactic disc scale radius
  yields expected counts ranging from $\sim 5$ to $\sim18$.
\item[{\it iii- Quasars:}] The number of expected quasars per square degree is
negligible in this comparison: If we extrapolate the results by
\citet{croom04}, and assume an order-of-magnitude colour transformation
B-K$_s$$\sim$4 for $z\lsim2.5$ objects \citep[see, for instance,][]{hewett06}, 
we expect $\sim 0.3$ quasars in our field, that is, $100\times$ smaller than 
the number of galaxies. Relaxing the B-K$_s$ assumption to $3<$ B-K$_s$ 
$<5$, the expected quasar counts in our frame would range between $\sim0.2$ 
and $\sim0.5$.
\end{description}

We thus conclude that, on the basis of simple statistics, $\sim25$ out of
the 47 sources detected in our K$_s$ image are possibly field galaxies at
{\it any} redshift, while 11 are expected to be Galactic stars. The
remaining 11 sources can be accounted for in terms of cosmic variance or
relaxing some of the assumptions we make in the estimates of field source
counts. As a simple estimate, assuming a Poissonian distribution for
  the expected number of observable field sources (stars+galaxies), we
  expect $\pm$12 sources within a 2-$\sigma$ deviation. Given those
  uncertainties our estimate of the field sources is consistent with the
  number of sources detected in our K$_s$ image.  Nevertheless, hereafter
we will consider the eventuality that they represent the tip of the iceberg
of a galaxy over-density associated to the S0927 system.

\subsection{Expectations for a galaxy cluster at $z\approx0.7$}\label{sec_cluster}

Here we evaluate whether the observed counts are consistent with the 
expectations for a cluster at $z\approx0.7$. The knee of the K$_s$-band 
luminosity function of galaxies at this redshift \citep[e.g.][]{cirasuolo08} 
corresponds to an apparent magnitude $m_{\rm Ks}\sim 18.5$, that is, one
magnitude brighter than our detection limit. 
Firstly we will compare our results to two galaxy clusters in the
nearby Universe studied in great detail. Then, we will step to high
redshift and consider what is actually observed in 
known $z\sim0.7$ clusters. 

We use the Virgo and Coma clusters as low-$z$ comparison terms. We 
emphasize that the velocity
dispersion of galaxies in Virgo is only $\sim 600$ km s$^{-1}$ 
\citep[as derived from the GOLDMine Database, see][]{goldmine}, while
it is $1080$ km s$^{-1}$ in Coma \citep{colless96}. In 
comparison the measured velocity of S0927 ($\approx 2600$ km s$^{-1}$) 
implies a considerably larger cluster mass, in the assumption that the
the shift in the emission lines reflects the cluster velocity dispersion.
Nevertheless, Virgo and Coma provide conservative
lower limits to the number of galaxies that should be observed in the S0927
field. Moreover, it will also allow us to take advantage of the wealth of
information available for low-$z$ galaxies. In particular, we will refer to
the GOLDMine Database \citep{goldmine} for multi-band photometry, to
\citet{virgostructure} for the 3D structure of the Virgo Cluster, and to
\citet{eisenhardt07} for Coma\footnote{We
will neglect the effects of cluster structure evolution, since this
exceeds by far the degree of accuracy we aim at in this work.}.

In order to shift Virgo and Coma to $z=0.7$, the following corrections 
are applied:
\begin{description}
\item[{\bf Distances and angular scales}] --  The Distance Moduli of Virgo
  and Coma are $31.2$ and $35.1$ mag, assuming luminosity distances of 17 and 
  96 Mpc respectively \citep{virgostructure,goldmine}. Therefore, each galaxy
  moved to $z=0.7$ should appear $12.0$ and $8.1$ mag fainter (without 
  considering any colour and filter correction: see below). The $3.6\times2.6$ 
  square arcmin field of our study corresponds to $\sim5.2\times3.8$ square 
  degrees at the actual distance of the Virgo Cluster, and to $\sim
  55\times40$ square arcmin for Coma. We dropped from our 
  analysis all the galaxies lying outside of boxes with these
  widths centered on M87 and NGC 4889, where the galaxy density 
  is the highest.
\item[{\bf Colour and filter corrections}] -- In order to minimize the
  effects of filter and $k$-corrections, we convert the rest-frame J 
  magnitudes into observed K$_s$
  magnitudes assuming the galaxy E--Sc templates presented in
  \citet{mannucci01}. The coverage of GOLDMine and of \citet{eisenhardt07}
  photometry is practically complete in the bright side of the luminosity 
  function \citep{gavazzi00}.
\item[{\bf Rejuvenation of stellar populations}] -- The stellar population
  of galaxies grew old in the 6.3 Gyr from $z=0.7$ to $z=0$. A detailed
  correction for this effect is challenging, since the star formation
  history of galaxies is mostly unknown. For the sake of simplicity, we
  adopt the parameterization proposed by \citet{gavazzi02}, with the scale
  age of the star formation history changing according to the galaxy type
  and magnitude, following \citet{cortese08}. On the other hand, NIR
  emission is insensitive to young stellar populations, and fades out
  relatively slowly. All the rejuvenation tracks of interest imply
  corrections ranging between $0.5$ and $1.1$ magnitudes.
\end{description}

\begin{figure}
\begin{center}
\includegraphics[width=0.49\textwidth]{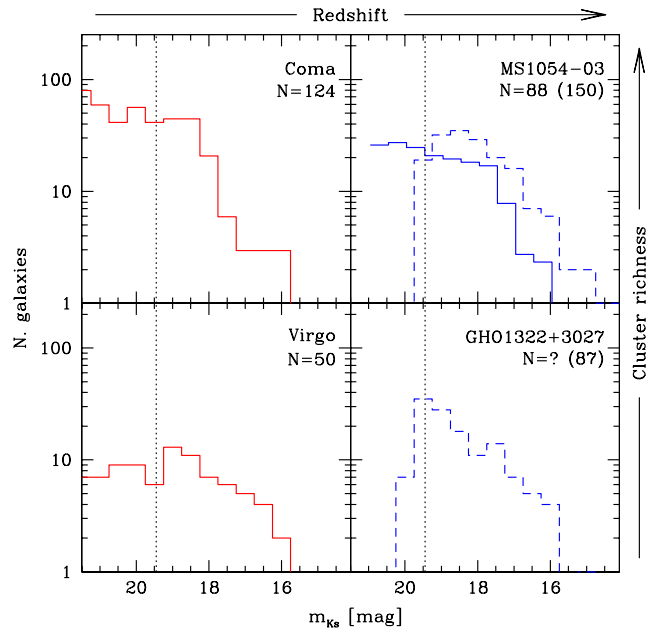}\\
\caption{The expected K$_s$-band  galaxy counts of the Virgo, Coma,
GHO 1322+3027 and MS 1054-03 Clusters, once ``moved'' to $z=0.7$ as 
described in the text. We note that we sampled different regimes
of cluster richness both at low and high-redshift.
The flux limit of our data is marked with a 
vertical dotted line. Solid lines refer only to those galaxies
belonging to the clusters, while dashed lines are not corrected for 
the contaminations of foreground sources. The number of objects exceeding
our sensitivity limits are also given in the case the correction for
foreground sources is (is not) applied.
}\label{fig_virgo}
\end{center}
\end{figure} 

The resulting luminosity functions are plotted in Figure \ref{fig_virgo},
\emph{left}.
We adopt the same flux cut as in our data, assuming that all galaxies
are practically point-like sources at $z=0.7$. We find that 50 sources 
are expected to be detected in our K$_s$ image, assuming that we are 
pointing at the core of a Virgo cluster twin centered on S0927,  
and 124 in the case of Coma. 
Such high numbers of galaxies are in sharp constrast with respect
to the 11 `extra' sources we observe, after star and field galaxy
subtraction (see \S \ref{field_stuff}). 

We cannot extend this comparison to other bands because of the 
poor statistics resulting from the relative shallowness of SDSS
photometry. For instance, we can derive the $z$-band expected 
luminosity function from GOLDMine V-band imaging of Virgo galaxies, 
and applying the same analysis described above. Nevertheless, once the 
$m_z<20.07$ flux limit is applied, only 8 sources are left. Such small 
numbers make this comparison inconclusive.

We now consider, as an additional check, the properties of observed galaxy
  clusters at $z\approx0.7$.  We will focus on the photometric catalogue by
  \citet{stanford02}. Among the $z\sim0.7$ clusters in their analysis, we
  select GHO 1322+3027 ($z=0.755$, $L_{\rm X}=0.14\times 10^{45}$ erg/s) and
  MS 1054-03 ($z=0.8231$, $L_{\rm X}=3.37\times 10^{45}$ erg/s) in order to
  probe two different regimes of cluster richness, as suggested by their
  X-ray luminosities\footnote{The estimates of $L_{\rm X}$ are taken from
    \citet{wu99} and \citet{jeltema01} for GHO 1322+3027 and MS1054-03,
    respectively.}. In particular, \citet{tran99} reported that the stellar
  velocity dispersion for MS1054-03 is $\sim 1200$ km/s, i.e. roughly
  half of the velocity difference observed between the red-NEL and the blue
  line systems in S0927.

In Figure \ref{fig_virgo}, \emph{right}, we plot the number of objects
  observed in GHO 1322+3027 and MS 1054-03 as a function of their K$_s$-band
  magnitude. In order to allow a direct comparison with our results, we
  again shift the distributions of $0.2$ and $0.4$ mag respectively, to
  take into account the different distance moduli with respect to
  S0927. Filter corrections from the observations of \citet{stanford02} and
  ours and stellar population aging effects are negligible. Up to 87 and 150
  sources exceeding our sensitivity limit are expected in these fields, that
  is, a factor 2 or 3 more than the \emph{total} observed sources in our
  K$_s$-band image of the S0927 field.  Concerning MS 1054-03,
  \citet{forster06} isolated the contribution of galaxies residing in the
  redshift range of the cluster (thus excluding most contaminating
  foreground and background objects). From their analysis, we infer that if
  S0927 were hosted in a cluster of galaxies similar to MS 1054-03, $\sim
  90$ galaxies with $z\approx0.7$ and $m_{\rm Ks}<19.45$ would be expected:
  again, many more than the observed.

We remark that, if we consider a lower threshold in the source detection,
e.g., we consider all objects with fluxes larger than 2 times the sky count
RMS in the PSF area, the ``extra'' sources would be 17 instead of 11. Such
a number is still 3 times below the predictions for a cluster such
as Virgo, and 5 (7) times fewer than those expected for a cluster similar to 
MS 1054-03 (Coma).

\section[]{Discussion and conclusions}

In this paper we present a deep K$_s$-band image of the $3.6\times2.6$
square arcmin field around S0927. We report the detection of 47 additional
sources with $m_{\rm K_s}<19.45$. These sources slightly exceed the
expectations from general field galaxy and star counts, but are much too few
to be consistent with the presence of a galaxy cluster in the field of
view.

We match the NIR information with archive GALEX FUV-NUV and SDSS
$ugriz$ photometry, and we confirm that 6-11 sources are consistent
with galaxies at $z\approx0.7$. Nevertheless, available optical
imaging is so shallow that our colour-based criteria can be applied
only to 32/77 (42 per cent) of the sample.

Using Virgo, Coma and the high-$z$ clusters GHO 1322+3027 and MS1054-03
  for comparison, we find that the sources observed in our K$_s$ image are
  several times fewer than expected if a similar cluster surrounded S0927.
The velocity dispersions of galaxies in Virgo, Coma and MS 1054-03 are
  $\sim 600$, $1080$ and $\sim1200$ km/s, while the redshift differences of
the line systems of S0927 correspond to $2650$ km/s, that is, at least
  a 2-$\sigma$ deviation. If the alleged cluster surrounding S0927 would be
richer (so that its velocity dispersion would be higher), the expected
galaxy counts will be even higher, which increases the disagreement with the
present analysis\footnote{Note that typical core radii of galaxy clusters
  are $\lsim0.3$ Mpc \citep{bahcall78}, therefore our image covered most of
  the alleged cluster independently of its size.}.

Summarizing, our analysis strongly disfavours the galaxy cluster
interpretation of S0927. As already suggested in \citet{heckman09}, 
future deep X-ray imaging of the field will provide an alternative independent 
check for the presence of a massive galaxy cluster in the S0927 field.
This may be especially important to probe clusters with extremely low luminous
to dark (dark matter + hot gas + diffuse light) mass ratios.

\section*{Acknowledgments}

We are grateful to the anonymous referee for his/her comments and
suggestions which substantially improved the paper quality. 
We thank Francesco Haardt, Ruben Salvaterra and Marco Scodeggio 
for fruitful discussions.
It is a pleasure to acknowledge the excellent support from the MDM
observatory staff. TIFKAM was funded by The Ohio State University, the MDM
consortium, MIT, and NSF grant AST-9605012. The HAWAII-1R array upgrade for
TIFKAM was funded by NSF Grant AST-0079523 to Dartmouth College. 
This research has made use of the NASA/IPAC Extragalactic Database 
(NED) which is operated by the Jet Propulsion Laboratory, California 
Institute of Technology, under contract with the National Aeronautics 
and Space Administration. This publication makes use of data
products from the Two Micron All Sky Survey, which is a joint
project of the University of Massachusetts and the Infrared
Processing and Analysis Center/ California Institute of Technology,
funded by the National Aeronautics and Space Administration and the
National Science Foundation.

\label{lastpage}
\onecolumn

\begin{center}
\begin{longtable}{ccccccccccc}
\caption[]{
{List of the sources detected in at least one band in this analysis.} 
(1) source ID; (2--3) RA and dec (J2000); (4--5) GALEX FUV and NUV magnitudes; 
(6--10) SDSS $ugriz$ magnitudes; (11) K$_s$-band magnitudes.} \label{tab_allsources}\\
\hline \\[-2ex]                                                     
  \multicolumn{1}{c}{ID} &  \multicolumn{1}{c}{RA(J2000)}
  &\multicolumn{1}{c}{Dec(J2000)} &\multicolumn{1}{c}{FUV}
  &\multicolumn{1}{c}{NUV} &\multicolumn{1}{c}{$u$} &\multicolumn{1}{c}{$g$}
  &\multicolumn{1}{c}{$r$} &\multicolumn{1}{c}{$i$} &\multicolumn{1}{c}{$z$}
  & \multicolumn{1}{c}{K$_s$} \\[0.5ex] 
\multicolumn{1}{c}{} & \multicolumn{1}{c}{} &\multicolumn{1}{c}{}
&\multicolumn{1}{c}{[mag]} &\multicolumn{1}{c}{[mag]} &\multicolumn{1}{c}{[mag]} &\multicolumn{1}{c}{[mag]} &\multicolumn{1}{c}{[mag]} &\multicolumn{1}{c}{[mag]} &\multicolumn{1}{c}{[mag]} &\multicolumn{1}{c}{[mag]}   \\[0.5ex]
\hline \\[-1.8ex]
\endfirsthead
\multicolumn{11}{c}{{\tablename} \thetable{} -- Continued} \\[0.5ex]
  \hline  \\[-2ex]
  \multicolumn{1}{c}{ID} &\multicolumn{1}{c}{RA(J2000)} &\multicolumn{1}{c}{Dec(J2000)} &\multicolumn{1}{c}{FUV} &\multicolumn{1}{c}{NUV} &\multicolumn{1}{c}{$u$} &\multicolumn{1}{c}{$g$} &\multicolumn{1}{c}{$r$} &\multicolumn{1}{c}{$i$} &\multicolumn{1}{c}{$z$} &\multicolumn{1}{c}{K$_s$}\\[0.5ex] 
\multicolumn{1}{c}{} & \multicolumn{1}{c}{} &\multicolumn{1}{c}{}
&\multicolumn{1}{c}{[mag]} &\multicolumn{1}{c}{[mag]} &\multicolumn{1}{c}{[mag]} &\multicolumn{1}{c}{[mag]} &\multicolumn{1}{c}{[mag]} &\multicolumn{1}{c}{[mag]} &\multicolumn{1}{c}{[mag]} &\multicolumn{1}{c}{[mag]}   \\[0.5ex]
\hline  \\[-1.8ex]
\endhead
  \multicolumn{11}{l}{{Continued on Next Page\ldots}} \\
\endfoot
  \\[-1.8ex] \hline 
\endlastfoot
 1 & $09:27:11.9$ & $+29:43:12$ & $     $ & $     $ & $     $ & $     $ & $21.77$ & $20.44$ & $19.89$ & $16.67$ \\   
 2 & $09:27:11.5$ & $+29:43:27$ & $     $ & $     $ & $     $ & $     $ & $     $ & $21.07$ & $20.33$ & $17.12$ \\   
 4 & $09:27:10.9$ & $+29:42:58$ & $     $ & $     $ & $     $ & $     $ & $     $ & $21.20$ & $     $ & $17.75$ \\   
 5 & $09:27:11.9$ & $+29:43:30$ & $     $ & $     $ & $     $ & $     $ & $     $ & $     $ & $     $ & $19.48$ \\   
 7 & $09:27:12.8$ & $+29:43:40$ & $     $ & $     $ & $     $ & $     $ & $     $ & $     $ & $     $ & $19.17$ \\   
 8 & $09:27:10.5$ & $+29:43:34$ & $     $ & $     $ & $     $ & $     $ & $21.36$ & $21.06$ & $     $ & $18.36$ \\   
10 & $09:27:10.2$ & $+29:42:56$ & $     $ & $     $ & $     $ & $     $ & $21.90$ & $21.07$ & $     $ & $17.39$ \\   
11 & $09:27:10.3$ & $+29:43:14$ & $     $ & $     $ & $     $ & $     $ & $     $ & $     $ & $     $ & $18.42$ \\   
12 & $09:27:10.2$ & $+29:43:24$ & $     $ & $     $ & $     $ & $     $ & $     $ & $21.36$ & $     $ & $17.73$ \\   
13 & $09:27:09.6$ & $+29:43:03$ & $     $ & $     $ & $     $ & $     $ & $     $ & $21.48$ & $     $ & $17.81$ \\   
15 & $09:27:12.6$ & $+29:43:44$ & $19.45$ & $18.48$ & $18.69$ & $18.42$ & $18.40$ & $18.40$ & $18.34$ & $16.22$ \\   
16 & $09:27:13.8$ & $+29:43:35$ & $19.94$ & $19.22$ & $19.38$ & $19.18$ & $19.14$ & $19.01$ & $19.13$ & $16.42$ \\   
17 & $09:27:07.9$ & $+29:43:00$ & $     $ & $     $ & $     $ & $     $ & $     $ & $     $ & $     $ & $19.23$ \\   
18 & $09:27:09.0$ & $+29:43:33$ & $     $ & $     $ & $     $ & $     $ & $     $ & $21.56$ & $     $ & $19.45$ \\   
19 & $09:27:12.9$ & $+29:42:34$ & $     $ & $     $ & $     $ & $21.90$ & $21.01$ & $20.59$ & $     $ & $17.97$ \\   
20 & $09:27:13.3$ & $+29:42:27$ & $     $ & $     $ & $     $ & $22.22$ & $     $ & $20.83$ & $     $ & $18.14$ \\   
24 & $09:27:14.8$ & $+29:43:21$ & $     $ & $     $ & $     $ & $20.28$ & $19.35$ & $18.89$ & $18.82$ & $16.31$ \\   
26 & $09:27:09.8$ & $+29:43:34$ & $     $ & $     $ & $     $ & $     $ & $     $ & $21.15$ & $     $ & $17.29$ \\   
27 & $09:27:10.2$ & $+29:43:43$ & $     $ & $     $ & $     $ & $     $ & $     $ & $20.79$ & $20.40$ & $17.19$ \\   
28 & $09:27:08.0$ & $+29:43:28$ & $     $ & $     $ & $     $ & $     $ & $     $ & $     $ & $     $ & $18.11$ \\   
29 & $09:27:09.7$ & $+29:43:43$ & $     $ & $     $ & $     $ & $     $ & $     $ & $     $ & $     $ & $19.40$ \\   
31 & $09:27:12.3$ & $+29:44:13$ & $     $ & $     $ & $21.12$ & $20.92$ & $20.92$ & $21.10$ & $     $ & $     $ \\   
32 & $09:27:09.2$ & $+29:42:21$ & $     $ & $     $ & $     $ & $21.18$ & $19.92$ & $19.30$ & $19.24$ & $17.15$ \\   
33 & $09:27:09.7$ & $+29:44:21$ & $     $ & $     $ & $     $ & $21.67$ & $20.13$ & $19.32$ & $19.03$ & $17.51$ \\   
34 & $09:27:16.6$ & $+29:43:07$ & $     $ & $     $ & $     $ & $21.12$ & $19.65$ & $19.09$ & $18.86$ & $15.94$ \\   
35 & $09:27:16.8$ & $+29:43:09$ & $     $ & $     $ & $     $ & $21.89$ & $21.21$ & $20.37$ & $19.97$ & $18.29$ \\   
36 & $09:27:16.3$ & $+29:43:12$ & $     $ & $     $ & $     $ & $     $ & $     $ & $21.37$ & $     $ & $     $ \\   
39 & $09:27:07.3$ & $+29:44:03$ & $     $ & $     $ & $     $ & $21.17$ & $20.50$ & $20.08$ & $19.95$ & $18.34$ \\   
40 & $09:27:12.8$ & $+29:42:11$ & $20.33$ & $19.46$ & $19.73$ & $18.96$ & $18.62$ & $18.43$ & $18.49$ & $17.59$ \\   
46 & $09:27:16.9$ & $+29:42:27$ & $     $ & $     $ & $     $ & $21.85$ & $21.48$ & $21.42$ & $     $ & $18.85$ \\   
47 & $09:27:17.1$ & $+29:42:30$ & $     $ & $     $ & $     $ & $     $ & $     $ & $     $ & $     $ & $19.03$ \\   
48 & $09:27:16.5$ & $+29:42:48$ & $     $ & $     $ & $     $ & $     $ & $21.82$ & $21.02$ & $     $ & $19.37$ \\   
49 & $09:27:16.9$ & $+29:42:44$ & $     $ & $     $ & $     $ & $     $ & $21.57$ & $21.02$ & $     $ & $18.77$ \\   
51 & $09:27:16.0$ & $+29:43:45$ & $     $ & $     $ & $     $ & $21.95$ & $21.06$ & $20.79$ & $     $ & $18.91$ \\   
52 & $09:27:16.1$ & $+29:44:08$ & $     $ & $     $ & $     $ & $21.58$ & $21.03$ & $20.80$ & $     $ & $19.59$ \\   
53 & $09:27:06.0$ & $+29:43:31$ & $     $ & $     $ & $     $ & $     $ & $     $ & $21.43$ & $     $ & $18.28$ \\   
55 & $09:27:12.4$ & $+29:42:17$ & $     $ & $     $ & $     $ & $     $ & $     $ & $     $ & $     $ & $17.67$ \\   
56 & $09:27:14.0$ & $+29:44:20$ & $     $ & $     $ & $     $ & $     $ & $21.13$ & $20.44$ & $20.48$ & $17.49$ \\   
57 & $09:27:15.0$ & $+29:44:03$ & $     $ & $     $ & $     $ & $     $ & $21.63$ & $20.93$ & $     $ & $18.04$ \\   
58 & $09:27:14.3$ & $+29:44:05$ & $     $ & $     $ & $     $ & $     $ & $     $ & $     $ & $     $ & $19.05$ \\   
61 & $09:27:17.5$ & $+29:43:43$ & $     $ & $     $ & $17.64$ & $16.15$ & $15.58$ & $15.36$ & $15.43$ & $14.00$ \\   
62 & $09:27:04.8$ & $+29:43:07$ & $     $ & $     $ & $     $ & $21.00$ & $19.78$ & $19.27$ & $19.09$ & $16.76$ \\   
63 & $09:27:04.5$ & $+29:44:01$ & $     $ & $     $ & $     $ & $20.69$ & $19.81$ & $19.33$ & $19.09$ & $18.21$ \\   
64 & $09:27:18.1$ & $+29:42:35$ & $     $ & $     $ & $     $ & $21.96$ & $20.68$ & $19.95$ & $19.67$ & $     $ \\   
65 & $09:27:18.5$ & $+29:43:18$ & $     $ & $     $ & $     $ & $     $ & $20.90$ & $19.47$ & $18.75$ & $17.19$ \\   
66 & $09:27:17.4$ & $+29:44:09$ & $     $ & $     $ & $     $ & $     $ & $     $ & $20.44$ & $19.70$ & $18.09$ \\   
67 & $09:27:18.8$ & $+29:42:55$ & $     $ & $     $ & $     $ & $21.44$ & $21.15$ & $20.81$ & $     $ & $     $ \\   
68 & $09:27:05.5$ & $+29:43:28$ & $     $ & $     $ & $     $ & $22.23$ & $21.09$ & $20.60$ & $20.29$ & $18.84$ \\   
70 & $09:27:04.4$ & $+29:42:42$ & $     $ & $     $ & $     $ & $     $ & $     $ & $21.00$ & $     $ & $19.08$ \\   
73 & $09:27:18.2$ & $+29:43:07$ & $     $ & $     $ & $     $ & $     $ & $     $ & $21.49$ & $     $ & $18.28$ \\   
74 & $09:27:18.4$ & $+29:43:09$ & $     $ & $     $ & $     $ & $     $ & $     $ & $21.34$ & $     $ & $     $ \\   
75 & $09:27:05.1$ & $+29:43:08$ & $     $ & $     $ & $     $ & $21.29$ & $20.84$ & $20.44$ & $20.16$ & $     $ \\   
76 & $09:27:12.8$ & $+29:43:48$ & $     $ & $     $ & $     $ & $     $ & $     $ & $21.26$ & $     $ & $17.47$ \\   
77 & $09:27:13.8$ & $+29:43:33$ & $     $ & $     $ & $     $ & $     $ & $     $ & $     $ & $     $ & $17.80$ \\   
\end{longtable}
\end{center}

\begin{center}
\begin{longtable}{@{}cccccccccc@{}}
\caption[]{
  Classification of the sources according to our colour-colour
  diagrams. (1) source ID; (2--5) source colours; (6--9) is the colour
  constraint satisfied? (see text for details) (10) source
  classification.}\label{tab_classif}\\
\hline \\[-2ex]                                 
ID  & $g$-$r$ & $r$-$i$ & $i$-K$_s$  & $z$-K$_s$  & $g$-$r$ & $r$-$i$ & $i$-K$_s$  & $z$-K$_s$   &  Source \\
    &  [mag]  &  [mag]  &  [mag]  &  [mag]  & $>1.2$? & $>0.8$? & $>2.5$? & $>2.0$? & classification  \\
\hline                                                                          
  1 & $     $ & $ 1.33$ & $ 3.77$ & $ 3.22$ &   & Y & Y & Y &   $z\approx0.7$ Gal   \\
  2 & $     $ & $     $ & $ 3.95$ & $ 3.21$ &   &   & Y & Y &   $z\approx0.7$ Gal   \\
  8 & $     $ & $ 0.30$ & $ 2.70$ & $     $ &   & N & Y &   &   Star?               \\
 10 & $     $ & $ 0.83$ & $ 3.68$ & $     $ &   & Y & Y &   &   $z\approx0.7$ Gal   \\
 15 & $ 0.02$ & $ 0.00$ & $ 2.18$ & $ 2.12$ & N & N & N & Y &   $z\approx0.7$ QSO   \\
 16 & $ 0.04$ & $ 0.13$ & $ 2.59$ & $ 2.71$ & N & N & Y & Y &   $z\approx0.7$ QSO   \\
 19 & $ 0.89$ & $ 0.42$ & $ 2.62$ & $     $ & N & N & Y &   &   Star?               \\
 24 & $ 0.93$ & $ 0.46$ & $ 2.58$ & $ 2.51$ & N & N & Y & Y &   low-$z$ Gal         \\
 27 & $     $ & $     $ & $ 3.60$ & $ 3.21$ &   &   & Y & Y &   $z\approx0.7$ Gal   \\
 31 & $ 0.00$ & $-0.18$ & $     $ & $     $ & N & N &   &   &   Star                \\
 32 & $ 1.26$ & $ 0.62$ & $ 2.15$ & $ 2.09$ & Y & N & N & Y &   $z\approx0.7$ Gal?  \\
 33 & $ 1.54$ & $ 0.81$ & $ 1.81$ & $ 1.52$ & Y & Y & N & N &   $z\approx0.7$ Gal?  \\
 34 & $ 1.47$ & $ 0.56$ & $ 3.15$ & $ 2.92$ & Y & N & Y & Y &   $z\approx0.7$ Gal?  \\
 35 & $ 0.68$ & $ 0.84$ & $ 2.08$ & $ 1.68$ & N & Y & N & N &   Star                \\
 39 & $ 0.67$ & $ 0.42$ & $ 1.74$ & $ 1.61$ & N & N & N & N &   Star                \\
 40 & $ 0.34$ & $ 0.19$ & $ 0.84$ & $ 0.90$ & N & N & N & N &   low-$z$ Gal         \\
 46 & $ 0.37$ & $ 0.06$ & $ 2.57$ & $     $ & N & N & Y &   &   Star                \\
 48 & $     $ & $ 0.80$ & $ 1.65$ & $     $ &   & N & N &   &   Star                \\
 49 & $     $ & $ 0.55$ & $ 2.25$ & $     $ &   & N & N &   &   Star                \\
 51 & $ 0.89$ & $ 0.27$ & $ 1.88$ & $     $ & N & N & N &   &   Star                \\
 52 & $ 0.55$ & $ 0.23$ & $ 1.21$ & $     $ & N & N & N &   &   Star                \\
 56 & $     $ & $ 0.69$ & $ 2.95$ & $ 2.99$ &   & N & Y & Y &   $z\approx0.7$ Gal?  \\
 57 & $     $ & $ 0.70$ & $ 2.89$ & $     $ &   & N & Y &   &   Star?               \\
 61 & $ 0.57$ & $ 0.22$ & $ 1.36$ & $ 1.43$ & N & N & N & N &   Star                \\
 62 & $ 1.22$ & $ 0.51$ & $ 2.51$ & $ 2.33$ & Y & N & Y & Y &   $z\approx0.7$ Gal?  \\
 63 & $ 0.88$ & $ 0.48$ & $ 1.12$ & $ 0.88$ & N & N & N & N &   Star                \\
 64 & $ 1.28$ & $ 0.73$ & $     $ & $     $ & Y & N &   &   &   Star?               \\
 65 & $     $ & $ 1.43$ & $ 2.28$ & $ 1.56$ &   & Y & N & N &   Star?               \\
 66 & $     $ & $     $ & $ 2.35$ & $ 1.61$ &   &   & N & N &   Star                \\
 67 & $ 0.29$ & $ 0.34$ & $     $ & $     $ & N & N &   &   &   Star                \\
 68 & $ 1.14$ & $ 0.49$ & $ 1.76$ & $ 1.45$ & N & N & N & N &   Star                \\
 75 & $ 0.45$ & $ 0.40$ & $     $ & $     $ & N & N &   &   &   Star                \\
\hline
\end{longtable}
\end{center}


\begin{thebibliography}{99}
\bibitem[\protect\citeauthoryear{Adelman-McCarthy et al.}{2008}]{adelman08}
  Adelman-McCarthy J.K., Ag\"ueros M.A., Allam S.S., Allende P.C., Anderson
  K.S.J., Anderson S.F., Annis J., Bahcall N.A. et al., 2008, ApJS, 175, 297
\bibitem[\protect\citeauthoryear{Bahcall}{1978}]{bahcall78} Bahcall N.A., 1975, ApJ, 198, 249
\bibitem[\protect\citeauthoryear{Bessell \& Brett}{1988}]{bessell88} Bessell
  M.S. \& Brett J.M., 1988, PASP, 100, 1134
\bibitem[\protect\citeauthoryear{Blanc et al.}{2008}]{blanc08} Blanc G.A.,
  Lira P., Barrientos L.F., Aguirre P., Francke H., Taylor E.N., Quadri R.,
  Marchesini D. et al., 2008, ApJ, 681, 1099 
\bibitem[\protect\citeauthoryear{Bogdanovic, Eracleous \&
    Sigurdsson}{2009}]{bogdanovic09} Bogdanovic T., Eracleous M., Sigurdsson
  S., 2009, arXiv:0809.326 2
\bibitem[\protect\citeauthoryear{Campanelli et al.}{2007}]{campanelli07}
  Campanelli M., Lousto C., Zlochower Y., Merritt D., 2007, ApJ, 659, L5 
\bibitem[\protect\citeauthoryear{Cirasuolo et al.}{2008}]{cirasuolo08} 
Cirasuolo M., McLure R.J., Dunlop J.S., Almaini O., Foucaud S., Simpson C., 2008, submitted to MNRAS (arXiv:0804.3471)  
\bibitem[\protect\citeauthoryear{Colless \& Dunn}{1996}]{colless96} Colless M. \& Dunn A.M., 1996, ApJ, 458, 435
\bibitem[\protect\citeauthoryear{Cortese et al.}{2008}]{cortese08} Cortese
  L., Boselli A., Franzetti P., Decarli R., Gavazzi G., Boissier S., Buat
  V., 2008, MNRAS, 386, 1157 
\bibitem[\protect\citeauthoryear{Croom et al.}{2004}]{croom04} Croom S.M.,
  Schade D., Boyle B.J., Shanks T., Miller L., Smith R.J., 2004, ApJ, 606,
  126 
\bibitem[\protect\citeauthoryear{Csabai et al.}{2003}]{csabai03} Csabai I.,
  Budav\'ari T., Connolly A.J., Szalay A.S., Gy\~ory Z., Ben\'{i}tez N.,
  Annis J., Brinkmann J. et al., 2003, AJ, 125, 580 
\bibitem[\protect\citeauthoryear{Dotti et al.}{2009}]{dotti09} Dotti M.,
  Montuori C., Decarli R., Volonteri M., Colpi M., Haardt F., 2009,
  arXiv:0809.3446 
\bibitem[\protect\citeauthoryear{Eisenhardt et al.}{2007}]{eisenhardt07} 
  Eisenhardt P.R., De Propris R., Gonzalez A.H., Stanford S.A., 
  Dickinson M., Wang M., 2007, ApJS, 169, 225
\bibitem[\protect\citeauthoryear{F\"{o}rster Schreiber et al.}{2006}]{forster06} F\"{o}rster Schreiber N.M., Franx M., Labb\'e I., Rudnick G., van Dokkum P.G., Illingworth G.D., Kuijken K., Moorwood A.F.M., Rix H.-W., R\"ottgering H., van der Werf P., 2006, AJ, 131, 1891
\bibitem[\protect\citeauthoryear{Francis et al.}{1991}]{francis91} Francis
  P.J., Hewett P.C., Foltz C.B. et al., 1991, ApJ, 373, 465 
\bibitem[\protect\citeauthoryear{Fukugita et al.}{2004}]{fukugita04}
  Fukugita M., Nakamura O., Schneider D.P., Doi M., Kashikawa N., 2004, ApJ,
  603, L65 
\bibitem[\protect\citeauthoryear{Gavazzi et al.}{1999}]{virgostructure}
  Gavazzi  G., Boselli A., Scodeggio M., Pierini D., Belsole E., 1999,
  MNRAS, 304, 595 
\bibitem[\protect\citeauthoryear{Gavazzi et al.}{2000}]{gavazzi00} Gavazzi
  G., Franzetti P., Scodeggio M., Boselli A., Pierini D., 2000, A\&A, 361,
  863 
\bibitem[\protect\citeauthoryear{Gavazzi et al.}{2002}]{gavazzi02} Gavazzi
  G., Bonfanti C., Sanvito G., Boselli A., Scodeggio M., 2002, ApJ, 576, 135 
\bibitem[\protect\citeauthoryear{Gavazzi et al.}{2003}]{goldmine} Gavazzi
  G., Boselli A., Donati A., Franzetti P., Scodeggio M., 2003, A\&A, 400,
  451 
\bibitem[\protect\citeauthoryear{Girardi et al.}{2005}]{girardi05} Girardi
  L., Groenewegen M.A.T., Hatziminaoglou E., da Costa L., 2005, A\&A, 436,
  895 
\bibitem[\protect\citeauthoryear{Heckman et al.}{2009}]{heckman09} Heckman
  T.M., Krolik J.H., Moran S.M., Schnittman J., Gezari S., 2009,
  arXiv:0810.1244 
\bibitem[\protect\citeauthoryear{Hewett et al.}{2006}]{hewett06} Hewett
  P.C., Warren S.J., Leggett S.K., Hodgkin S.T., 2006, MNRAS, 367, 454 
\bibitem[\protect\citeauthoryear{Jeltema et al.}{2001}]{jeltema01} Jeltema T.E., 
  Canizares C.R., Bautz M.W., Malm M.R., Donahue M., Garmire G.P., 2001, ApJ, 562, 124 
\bibitem[\protect\citeauthoryear{Komossa et al.}{2008}]{komossa08} Komossa
  S., Zhou H., Lu H., 2008, ApJ, 678, L81 
\bibitem[\protect\citeauthoryear{Mannucci et al.}{2001}]{mannucci01}
  Mannucci F., Basile F., Poggianti B.M., Cimatti A., Daddi E., Pozzetti L.,
  Vanzi L., 2001, MNRAS, 326, 745 
\bibitem[\protect\citeauthoryear{Morrissey et al.}{2005}]{morrissey05}
  Morrissey P., Schiminovich D., Barlow T.A. et al., 2005, ApJ, 619, L7 
\bibitem[\protect\citeauthoryear{Niemack et al.}{2009}]{niemack09} Niemack
  et al.,  2009, ApJ, 690, 89 
\bibitem[\protect\citeauthoryear{Pogge et al.}{1998}]{pogge98} Pogge R.W.,
  Depoy D.L., Atwood B., O'Brien T.P., Byard P.L., Martini P.,Stephens A.,
  Gatley I., et al., 1998, SPIE, 3354, 414  
\bibitem[\protect\citeauthoryear{Shields, Bonning \&
    Salviander}{2009}]{shields09} Shields G.A., Bonning E.W., Salviander S.,
  2008, arXiv:0810.2563 
\bibitem[\protect\citeauthoryear{Schnittman \&
    Buonanno}{2007}]{schnittman07} Schnittman J.D. \& Buonanno A., 2007,
  ApJ, 662, L63 
\bibitem[\protect\citeauthoryear{Stanford et al.}{2002}]{stanford02} Stanford 
  S.A., Eisenhardt P.R., Dickinson M., Holden B.P., De Propris R., 2002,
  ApJS, 142, 153
\bibitem[\protect\citeauthoryear{Tran et al.}{1999}]{tran99} Tran K.V.H., Kelson D.D., van Dokkum P., Franx M., Illingworth G.D., Magee D., 1999, ApJ, 522, 39
\bibitem[\protect\citeauthoryear{Wu Xue \& Fang}{1999}]{wu99} Wu X., Xue Y., \& Fang L., 1999, ApJ, 524, 22
\end{thebibliography}
\end{document}